\documentclass{article}

\usepackage{graphicx,latexsym,amssymb,amsmath,stmaryrd,epsfig}

\newcommand\var[1]{\langle #1\rangle}
\def\ra{\rightarrow}
\def\lra{\longrightarrow}
\def\nulll{\mbox{\it Null\/}}
\def\pred{\mbox{\it Pred\/}}
\def\fa{\forall}
\def\ex{\exists}

\newbox\tempa
\newbox\tempb
\newdimen\tempc
\def\mud#1{\hfil $\displaystyle{\mathstrut #1}$\hfil}
\def\rig#1{\hfil $\displaystyle{#1}$}
\def\irulehelp#1#2#3{\setbox\tempa=\hbox{$\displaystyle{\mathstrut #2}$}%
                        \setbox\tempb=\vbox{\halign{##\cr
        \mud{#1}\cr
        \noalign{\vskip\the\lineskip}%
        \noalign{\hrule height 0pt}%
        \rig{\vbox to 0pt{\vss\hbox to 0pt{${\; #3}$\hss}\vss}}\cr
        \noalign{\hrule}%
        \noalign{\vskip\the\lineskip}%
        \mud{\copy\tempa}\cr}}%
                      \tempc=\wd\tempb
                      \advance\tempc by \wd\tempa
                      \divide\tempc by 2 }
\def\irule#1#2#3{{\irulehelp{#1}{#2}{#3}%
                     \hbox to \wd\tempa{\hss \box\tempb \hss}}}

\newtheorem{definition}{Definition}[section]
\newtheorem{proposition}{Proposition}[section]
\newtheorem{theorem}{Theorem}[section]

\newenvironment{example}{\medskip \noindent {\em Example.}~}

\newenvironment{proof}{\noindent {\em Proof.}~}

\title{On the Expressive Power of Schemes}

\author{Gilles Dowek\thanks{INRIA,
23 avenue d'Italie,
CS 81321, 75214 Paris Cedex 13, France,
{\tt \small gilles.dowek@inria.fr}}
~and Ying Jiang
\thanks{State Key Laboratory of Computer Science,
Institute of Software,
Chinese Academy of Sciences,
P.O. Box 8718, 100190 Beijing, China,
{\tt \small jy@ios.ac.cn}}}

\date{}

\begin{document}
\maketitle

\begin{abstract}
We present a calculus, called {\em the scheme-calculus}, that permits
to express natural deduction proofs in various theories. Unlike
$\lambda$-calculus, the syntax of this calculus sticks closely to the
syntax of proofs, in particular, no names are introduced for the hypotheses.
We show that despite its non-determinism, some typed scheme-calculi
have the same expressivity as the corresponding typed
$\lambda$-calculi.
\end{abstract}


\section{Introduction}

We present a calculus, called {\em the scheme-calculus}, that permits
to express natural deduction proofs without introducing 
names for the hypotheses.

\subsection{A scheme calculus}

In the algorithmic interpretation of proofs, introduced by Brouwer,
Heyting, and Kolmogorov, proofs are 
expressed by terms of a typed $\lambda$-calculus. In such a calculus, 
two kinds of variables are often used: those of the logic and
those introduced to name the hypotheses. In System $F$, for instance, 
type variables and proof variables are often
distinguished.  

When variables are introduced to name the hypotheses,
the two occurrences of the proposition $A$ in the context of the
sequent $A, A \vdash A$ must be distinguished, and thus the contexts must
be multisets of propositions. In contrast, in automated theorem proving,
in order to reduce the search space ({\em e.g.} to a finite space),
the contexts of the sequents are often considered as sets of propositions
\cite{Kleene}. Thus, slightly different notions of sequents are used
in proof-theory and in automated theorem proving.  Moreover, 
these hypothesis names make the proofs of a given proposition a 
non-context-free language, even in the minimal propositional logic
\cite{Zaionc,Joly,contfree}.

In this paper, we introduce a calculus, called the {\em
scheme-calculus}, that permits to express proofs without introducing names
for the hypotheses
and where the contexts are just sets of hypotheses. In other words, 
we keep the variables of predicate
logic, but do not introduce another category of variables for the
hypotheses.  

In the scheme-calculus, the proofs of a given proposition in minimal
propositional logic and even in the positive fragment of minimal
predicate logic form a context-free language. In fact,
this scheme-calculus stems from previous works on the
grammatical properties of sets of $\lambda$-terms
\cite{BenYelles,Zaionc1988,TakahashiAkamaHirokawa,ComonJurski,BrodaDamas,BrodaDamas2,bracket,contfree,Salvati}.

From the grammar generating the schemes of a given type, we can
build an algorithm generating all the $\lambda$-terms of this type, as
each scheme corresponds to a finite number of terms that can be
computed from it \cite{contfree}.  
A scheme
containing $n$ abstractions and $p$ variables aggregate up to $p^n$
$\lambda$-terms.  In this sense, 
more proofs are identified in the scheme-calculus 
than in the $\lambda$-calculus, but, 
unlike in the formalisms based on proof irrelevance,
not all the proofs are identified, for instance the terms 
$\lambda x_P \lambda y_P~x$ and $\lambda x_P \lambda y_P~y$ are
identified, but the terms 
$\lambda x_P \lambda f_{P \Rightarrow P}~x$ and 
$\lambda x_P \lambda f_{P \Rightarrow P}~(f~x)$ are not. 

Despite its simplicity, we show that this scheme-calculus
is as expressive as the dependently-typed $\lambda$-calculus: for some
type systems, all the functions that are provably total in
impredicative ({\em i.e.} second-order)
arithmetic can be expressed in the scheme-calculus.
In this expressivity result, the 
determinism does not come from a local property, such as confluence, as
for the $\lambda$-calculus, but from the subject-reduction property and
the fact that dependent types are powerful enough to specify the value of
terms.

\subsection{The notion of variable}

To understand the basic idea of the scheme-calculus, it is useful to
go back to the origin of the notion of variable. 
A term expressing a function is usually built using
a function-former, often written as $\lambda$, and a place-holder for the
yet unknown argument of the function, sometimes written as $\Box$.
For instance, the
function mapping a number to its double can be expressed by the term
$$\lambda (2 \times \Box)$$
Applying this term to $4$ yields a term that reduces to $2 \times 4$.

But, when applying the term
$$\lambda \lambda~(2 \times \Box \times \Box \times \Box)$$ that
contains several occurrences of the symbol $\lambda$, to the arguments
$4$ and $5$, for instance, we may get eight
different syntactic results by replacing each occurrence of the symbol $\Box$
either by $4$ or by $5$. Hence arises the need of
a pointer associating a function-former occurrence $\lambda$ to each
place-holder occurrence $\Box$.

In the $\lambda$-calculus, this pointer is expressed by giving a name to
each occurrence of a $\lambda$ and to each occurrence of a $\Box$.
The $\lambda$ associated to a place-holder $\Box_x$ is then the
first $\lambda_x$ above it in the term seen as a tree. This way, the
function mapping two numbers to the double of the product of the
square of the first and of the second is written as
$$\lambda_x \lambda_y~(2 \times \Box_x \times \Box_x \times \Box_y)$$
or, in a simpler way, as
$$\lambda x \lambda y~(2 \times x \times x \times y)$$

Other solutions have been investigated. A solution related to
Bourbaki's is to express the pointer with a directed edge from each $\Box$ to
the corresponding $\lambda$
\begin{center}
\epsfig{file=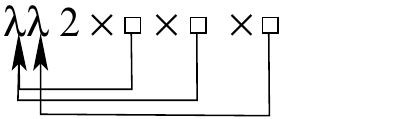}
\end{center}
While, in the solution proposed by de Bruijn, each $\Box$ is assigned the
height of its associated $\lambda$ above it. So we get
$$\lambda \lambda~(2 \times \Box_2 \times \Box_2 \times \Box_1)$$
Applying these three terms to the terms $4$ and $5$ yields terms that reduce,
in each formalism, to $2 \times 4 \times 4 \times 5$ only.

In many cases, both $\lambda$s and $\Box$s are typed and the pointers
must relate objects of the same type. 
This identity of types
guarantees the subject-reduction property: the reduction of a
well-typed term yields a term of the same type. 
Knowing the type of each
$\lambda$ and $\Box$ often reduces the possibilities of linking
occurrences of the symbols
$\Box$
and $\lambda$ in a raw term. For instance, in the raw term
$$\lambda_{scal} \lambda_{vect} (2 . \Box_{scal} . \Box_{scal} . \Box_{vect})$$
there is only one way to associate a $\lambda$ to each $\Box$, but
in the raw term
$$\lambda_{nat} \lambda_{nat}~(2 \times \Box_{nat} \times \Box_{nat} \times
\Box_{nat})$$ there are still eight ways to associate a $\lambda$ to
each $\Box$.

In the scheme-calculus, instead of distinguishing eight terms 
$\lambda x \lambda y~(2 \times x \times x \times x)$, $\lambda x
\lambda y~(2 \times x \times x \times y)$, ..., we consider a single
{\em scheme} $\lambda_{nat} \lambda_{nat}~(2
\times \var{nat} \times \var{nat} \times \var{nat})$, 
where $\var{nat}$ is the canonical ({\em i.e.} only) variable of type $nat$.
In this scheme, each place-holder is possibly associated
to any function-former above it, provided they have the same
type. 
The scheme 
$(\lambda_{nat} \lambda_{nat}~(2
\times \var{nat} \times \var{nat} \times \var{nat})~4~5)$
aggregates
eight terms
and reduces, in a
non-deterministic way, to $(2 \times 4 \times 4 \times 4)$,
$(2 \times 4 \times 4 \times 5)$, ...
The reduction of schemes is therefore non-deterministic, but
it does enjoy the subject-reduction property.

In general, the scheme
$((\lambda_A ... \lambda_A \var{A})~t_1~t_2~...~t_n)$ reduces, 
in a non-deterministic way, to each of the $t_i$s. This is typical of 
non-deterministic extensions of $\lambda$-calculus, such as 
G. Boudol's $\lambda$-calculus with multiplicities \cite{Boudol}, where 
the term $(\lambda x~x)~(t_1~|~...~|~t_n)$ reduces also in a 
non-deterministic way to each of $t_i$'s.

In the $\lambda$-calculus, when we apply the substitution 
$(f~y)/x$ to the term $ \lambda y:B~(g~x~y)$, we must rename the
bound variable $y$ in order to avoid the variable capture. 
As there is only one variable of type $B$ in the scheme calculus, 
we are no longer able to rename the variables this way and 
the variable captures cannot be avoided.

\subsection{The algorithmic interpretation of proofs}

In the algorithmic interpretation of
proofs, the subject-reduction property is more important
than the uniqueness of results. For instance, consider the natural deduction
proof
$$\irule{\irule{\irule{\irule{\irule{}
                                    {A, A \vdash A}
                                    {\mbox{ax}}
                             }
                             {A \vdash A \Rightarrow A}
                             {\mbox{$\Rightarrow$i}}
                      }
                      {\vdash A \Rightarrow A \Rightarrow A}
                      {\mbox{$\Rightarrow$i}}
                  ~~~~~~~~~~
                 \irule{t}
                     {\vdash A}
                     {}
               }
               {\vdash A \Rightarrow A}
               {\mbox{$\Rightarrow$e}}
         ~~~~~~~~~~~~~~~~~~~~~~~
         \irule{u}
               {\vdash A}
               {}
        }
        {\vdash A}
        {\mbox{$\Rightarrow$e}}$$
where $t$ and $u$ are two cut free proofs of the sequent $\vdash A$.
This proof can be reduced, in a non-deterministic way, to $t$ or to $u$, but
in both cases, we get a cut free proof of $\vdash A$.

When we associate a term of $\lambda$-calculus to this proof,
we must associate a variable name to each hypothesis of the
sequent $A, A \vdash A$, and we must choose the
variable used in the axiom rule. Different choices lead to different
proof-terms: $((\lambda \alpha \lambda \beta~\alpha)~t~u)$ and
$((\lambda \alpha \lambda \beta~\beta)~t~u)$, and each of these terms
reduces to a unique normal form.

This example shows that, in some presentations of natural deduction
with unnamed hypotheses, proof reduction is non-deterministic, and
$\lambda$-calculus introduces determinism in a somewhat artificial
way.  

As shown by Statman and Leivant (see
\cite{TroelstraSchwichtenberg,Leivant}) the proof reduction process
defined directly on natural deduction proofs with unnamed hypotheses
is not strongly normalizing, while that of $\lambda$-calculus is. This
non-termination can be seen as a consequence of the fact that variable
captures are allowed.  As, in general, termination is lost in the
scheme-calculus, a strategy must be chosen.

\subsection{Names and specifications}

In the cross-fertilization of the theories
of proof languages and of programming languages, the expression of natural
deduction proofs in $\lambda$-calculus can be seen as the importation
in proof theory of the concept of variable name, that is familiar in
the theory of programming languages. On the opposite, the
scheme-calculus can be seen as an importation in the theory of
programming languages of the concept of anonymous hypothesis, that is
familiar in proof theory.

Yet, this idea of anonymous resource is not completely new in
computer science. For instance, when we connect a computer to a local
network, we just need to use any unnamed Ethernet cable. Its type 
``Ethernet cable'' is
sufficient to guarantee the connection to the network.
In the same way, when a type system is strong enough to specify the value 
returned by 
a program, there is no need to give names to different programs of the same
type: when such a program is needed, any program, that has the right type,
goes. Identifying programs by their specification and not by their name
may be a way to avoid the proliferation of variable names in programs and
other formal objects.

\medskip

The main calculus we shall introduce in this paper is a scheme-calculus
with dependent types (Section \ref{secdep}), that permits to
express proofs
of various theories in minimal predicate logic. We shall prove three
properties of this
dependently-typed scheme-calculus, that are subject-reduction
(Section \ref{secsubred}), normalization (Section
\ref{secnorm}), and an expressivity result (Section
\ref{secexpress}). 
As an introductory example, we start with a
simply-typed scheme-calculus.

\section{A simply-typed scheme-calculus}
\label{secsimple}

\subsection{The calculus}

\begin{definition}[Simple types]
Let ${\cal P}$ be a set of atomic types. The {\em simple types} are
inductively defined by
$$A = P~|~A \Rightarrow A'$$
with $P \in {\cal P}$.
\end{definition}

\begin{definition}[Context]
A {\em context} is a finite set of simple types.
\end{definition}

\begin{definition}[Simply-typed schemes]
Schemes are inductively defined by
$$t = \var{A}~|~\lambda_A t~|~(t~t')$$
\end{definition}
The scheme $\var{A}$ is the canonical variable of type $A$,
$\lambda_A t$ is the scheme obtained by abstracting 
the canonical variable $\var{A}$ of type $A$
in $t$, and $(t~t')$ is the application of the scheme $t$ to the 
scheme $t'$.

The typing rules are given in Figure \ref{simpletyping}.
Notice that as contexts are sets, if $A$ is an element of $\Gamma$, then 
$\Gamma \cup \{A\}$ is just $\Gamma$.
For instance, using these rules, the scheme 
$\lambda_A \lambda_A \lambda_A \var{A}$ can be given the type 
$A \Rightarrow A \Rightarrow A \Rightarrow A$ with the following derivation.
$$\irule{\irule{\irule{\irule{}
                             {A \vdash \var{A}:A}
                             {} 
                      }
                      {A \vdash \lambda_A~\var{A}:A \Rightarrow A}
                      {}
               }
               {A \vdash \lambda_A \lambda_A~\var{A}:A \Rightarrow A \Rightarrow A}
               {}
        }
        {\vdash \lambda_A \lambda_A \lambda_A~\var{A}:A \Rightarrow A \Rightarrow A \Rightarrow A}
        {}$$

\begin{definition}[Scheme in context]
A {\em scheme in context} is a pair $t_{\Gamma}$ where $t$ is a scheme and
$\Gamma$ is a context such that $t$ is well-typed in $\Gamma$.
\end{definition}
We sometimes omit the context $\Gamma$ when there is no ambiguity.

\begin{figure}[t]
\noindent\framebox{\parbox{\columnwidth }{
$$\irule{}{\Gamma \vdash \var{A} :A}{A \in \Gamma}$$
$$\irule{\Gamma \cup \{A\} \vdash t : B}{\Gamma \vdash \lambda_A t : A \Rightarrow B}
{}$$
$$\irule{\Gamma \vdash t : A \Rightarrow B~~~\Gamma \vdash u: A}{\Gamma \vdash (t~u): B}{}$$
\caption{Simply-typed schemes\label{simpletyping}}}}
\end{figure}

\subsection{Reduction}

When reducing the underlined redex in the scheme
$$\lambda_A \ldots \lambda_A (\underline{\lambda_A (\lambda_A \ldots\lambda_A
\var{A})~u})$$
the variable $\var{A}$ may be bound by the $\lambda_A$ of the redex,
but it may also be bound by another $\lambda_A$, either higher or lower in the
scheme. So, in general, the scheme $u$ may be substituted for
the variable $\var{A}$ or not, hence the non-determinism of the substitution.

However, if the variable $\var{A}$ is bound neither higher nor lower 
in the term, the only possible binder for $\var{A}$ is that of the redex. 
In such a case, the variable $\var{A}$ must be substituted.
Thus, the fact that the scheme $u$ may or must be
substituted for the variable $\var{A}$ depends not only on the reduced redex
but also on the position of this redex in the scheme. Therefore,
the reduction relation cannot
be defined on schemes. Instead, it has to be defined on schemes in contexts.

To define the reduction relation, we must first set up a notion of
substitution.  A {\em substitution} is a function of finite domain,
written as $[t_1/A_1, ..., t_n/A_n]$, associating schemes $t_1, ..., t_n$
of types $A_1, ..., A_n$, respectively, to the variables $\var{A_1},
..., \var{A_n}$.  Applying a substitution to a scheme may produce
several results, thus this application produces a set of results.
Moreover, this application is always performed with respect to some
context $\Gamma$ that specifies the variables for which the
substitution may or must be performed. More precisely, when $A$ is in
$\Gamma$, we may choose whether we substitute the canonical variable
of type $A$ or not and when $A$ is not in $\Gamma$, this substitution is
forced.  If $\sigma$ is a substitution, $t$ is a scheme and $\Gamma$ a
context, we write $\sigma_{\Gamma} t$ for the result of the
application of $\sigma$ to $t$, with respect to the context $\Gamma$.

To simplify the notations, if $S$ and $S'$ are sets of schemes, we write
$\lambda_A S$ for the set of schemes of the form $\lambda_A t$ for
$t$ in $S$ and $(S~S')$ for the set of schemes of the form $(t~t')$
for $t$ in $S$ and $t'$ in $S'$.

\begin{definition}[Substitution]
\label{defsubstsimpl}
~
\begin{itemize}
\item $\sigma_{\Gamma} \var{A} = \{\var{A}, \sigma(A)\}$
if $A \in dom(\sigma)$ and $A \in \Gamma$,

\item $\sigma_{\Gamma} \var{A} = \{\sigma(A)\}$
if $A \in dom(\sigma)$ and $A \not\in \Gamma$,

\item $\sigma_{\Gamma} \var{A} = \{\var{A}\}$
if $A \not\in dom(\sigma)$,

\item $\sigma_{\Gamma}(\lambda_A t) = \lambda_A
\sigma_{(\Gamma \cup \{A\})} t$,

\item $\sigma_{\Gamma} (t~u) =
(\sigma_{\Gamma} t~\sigma_{\Gamma}u)$.
\end{itemize}
\end{definition}

\begin{definition}[Reduction]
\label{defredsimpl}
The {\em one step top level $\beta$-reduction}, written as
$\lra$, is defined by the rule
$$((\lambda_A t)~u)_{\Gamma} \lra v_{\Gamma}$$
where $A$, $t$, $u$, and $\Gamma$ are arbitrary and $v$ is any element of
$[u/A]_{\Gamma}t$.

The {\em one step $\beta$-reduction} relation
$\rhd$ is the contextual closure of the relation $\lra$.
It is inductively defined by

\begin{itemize}
\item if $t_{\Gamma} \lra t'_{\Gamma}$, then $t_{\Gamma} \rhd t'_{\Gamma}$,
\item if $t_{\Gamma} \rhd t'_{\Gamma}$, then
$(t~u)_{\Gamma} \rhd (t'~u)_{\Gamma}$,
\item if $u_{\Gamma} \rhd u'_{\Gamma}$, then
$(t~u)_{\Gamma} \rhd (t~u')_{\Gamma}$,
\item if $t_{\Gamma \cup \{A\}} \rhd t'_{\Gamma \cup \{A\}}$, then
$(\lambda_A~t)_{\Gamma} \rhd (\lambda_A~t')_{\Gamma}$.
\end{itemize}

This {\em $\beta$-reduction} relation $\rhd^*$ is the reflexive-transitive
closure of $\rhd$.
\end{definition}

The reduction relation is not confluent.
Indeed, if $A \not\in \Gamma$,
the scheme $((\lambda_A \lambda_A~\var{A})~t~u)_{\Gamma}$ reduces to both
$t_{\Gamma}$ and $u_{\Gamma}$, in a non-deterministic way. This formalizes
the intuition that, in the
scheme $(\lambda_A \lambda_A \var{A})_{\Gamma}$, the variable $\var{A}$ may be
bound by either of the occurrences of the symbol $\lambda_A$.

More surprisingly this reduction relation is not strongly normalizing.

\subsection{Counter-examples to strong normalization}
\label{counterex}

\begin{proposition}[Statman \cite{TroelstraSchwichtenberg}, 
Leivant \cite{Leivant}]
The simply-typed scheme-calculus is not strongly normalizing.
\end{proposition}

\proof{Let $t = ((\lambda_A \var{A})~\var{A})_{A}$.
The scheme $((\lambda_A t)~t)_{A}$ reduces to each of the elements of
$[t/A]_{A}t$, {\em e.g.} to $((\lambda_A t)~t)_{A}$.}

\medskip

This counter-example shows that, when we express natural deduction
with sequents without naming the hypotheses, proof reduction is not
strongly normalizing. For instance, the proof
$$\hspace*{-1.2cm}
\irule{\irule{\irule{\irule{\irule{}
                                    {A \vdash A}
                                    {\mbox{ax}}
                             }
                             {A \vdash A \Rightarrow A}
                             {\mbox{$\Rightarrow$i}}
                              ~~~~~~~
                             \irule{}
                                   {A \vdash A}
                                   {\mbox{ax}}
                      }
                      {A \vdash A}
                      {\mbox{$\Rightarrow$e}}
               }
               {A \vdash A \Rightarrow A}
               {\mbox{$\Rightarrow$i}}
               ~~~~~~~~~~~~~~~~~~~~~~~~~~~~~~~
               \irule{\irule{\irule{}
                                   {A \vdash A}
                                   {\mbox{ax}}
                            }
                            {A \vdash A \Rightarrow A}
                            {\mbox{$\Rightarrow$i}}
                            ~~~~~~~
                            \irule{}
                                  {A \vdash A}
                                  {\mbox{ax}}
                      }
                      {A \vdash A}
                      {\mbox{$\Rightarrow$e}}}
                      {A \vdash A}
                      {\mbox{$\Rightarrow$e}}$$
contains a cut. If we eliminate this cut, we have to replace in the proof $t$
$$\irule{\irule{\irule{}
                      {A \vdash A}
                      {\mbox{ax}}
               }
               {A \vdash A \Rightarrow A}
               {\mbox{$\Rightarrow$i}}
          ~~~~~~~
          \irule{}
                {A \vdash A}
                {\mbox{ax}}
         }
         {A \vdash A}
         {\mbox{$\Rightarrow$e}}$$
the axiom rules on the proposition $A$ with the proof $t$ itself.
As $A$ was already in the context, before being
introduced by the $\Rightarrow$i rule of the cut, we may choose
to replace each axiom rule or not. If we replace both, we get back the
proof we started with.

\medskip

This counter-example is based on 
the fact that the scheme-calculus permits the
substitution
of bound variables. Yet,
even if we forbid this substitution of bound variables, 
the variable captures of the scheme-calculus are sufficient to jeopardize
strong normalization. We give here another counter-example.

\begin{example}
Consider the context $\Gamma = \{A \Rightarrow B, B \Rightarrow A,
A \Rightarrow B \Rightarrow A, A, B\}$, and the schemes in $\Gamma$,
$f = \var{A \Rightarrow B}$, $g = \var{B \Rightarrow A}$,
$h = \var{A \Rightarrow B \Rightarrow A}$
$$a = ((\lambda_B \var{A}) (f~\var{A}))$$
$$b = ((\lambda_A \var{B}) (g~\var{B}))$$
$$u_0 = (h~a~b)$$
$$u_{n+1} = (g~(f~u_n))$$
$$v_n = ((\lambda_B~u_n)~(f~u_n))$$

Remark that, for each $i$, the schemes $a$ and $b$ are subschemes of the
scheme $u_i$ and that they do not occur in the scope of any binder.

The scheme $v_n$ reduces to
$[(f~u_n)/B]_{\Gamma} u_n$ that contains a subscheme
$[(f~u_n)/B]_{\Gamma}b$, {\em i.e.}
$((\lambda_A~(f~u_n))(g~(f~u_n)))$
that, in turn, reduces to
$[(g~(f~u_n))/A]_{\Gamma}(f~u_n)$, that contains a subscheme
$[(g~(f~u_n))/A]_{\Gamma} a$, {\em i.e.}
$((\lambda_B (g~(f~u_n))) (f~(g~(f~u_n))))$ that is $v_{n+1}$.
Therefore $v_n$ reduces to a scheme that contains $v_{n+1}$ as a subscheme.
\end{example}

\subsection{Strategies}

As with any non-deterministic system, we can restrict the reduction of
the scheme-calculus by defining strategies. In the
scheme-calculus, non-determinism arises from two different origins. First, as
in the $\lambda$-calculus, when a scheme contains several redex occurrences, we
may choose to reduce one or another first. Then, once the redex occurrence is
chosen, we still have several ways to reduce it, because substitution itself
is non-deterministic.

The simplest strategies are obtained by restricting the
non-determinism of the substitution.

The definition of the substitution of the {\em minimal strategy} is the same
as that of the general notion of substitution
(Definition \ref{defsubstsimpl}), except for the first clause:
here we take the scheme $\var{A}$ only, {\em i.e.}
\begin{itemize}
\item $\underline{\sigma}_{\Gamma}\var{A} = \{\var{A}\}$,
if $A \in dom(\sigma)$ and $A \in \Gamma$.
\end{itemize}
Notice that, in this case, $\underline{\sigma}_{\Gamma} t$ is always a
singleton.
Its only element also is denoted by
$\underline{\sigma}_{\Gamma}t$.

For instance, if $A \not\in \Gamma$, the scheme
$((\lambda_A \lambda_A \var{A})~t)_{\Gamma}$
reduces to $(\lambda_A \var{A})_{\Gamma}$, and so does the scheme
$(\lambda_A ((\lambda_A \var{A})~t))_{\Gamma}$.
But the scheme $((\lambda_A \var{A})~t)_{\Gamma}$ reduces to $t_{\Gamma}$.

The dual strategy is the {\em maximal strategy}.
The definition of the substitution of this strategy is the same as that
of the general notion of
substitution except for the first clause:
here we take the scheme $\sigma(A)$ only, {\em i.e.}
\begin{itemize}
\item $\overline{\sigma}_{\Gamma} \var{A} = \{\sigma(A)\}$, if $A \in dom(\sigma)$ and
$A \in \Gamma$.
\end{itemize}
The scheme $((\lambda_A \lambda_A \var{A})~t)_{\Gamma}$ now
reduces to $(\lambda_A t)_{\Gamma}$ and so does the scheme
$(\lambda_A ((\lambda_A \var{A})~t))_{\Gamma}$.

Intuitively, in the minimal strategy, we
substitute a variable if we have to, while in the maximal strategy, we
substitute a variable if we are able to.

A more complex strategy is the {\em strategy with reference to the closest
binder}, also known as the {\em total discharge strategy}
\cite{TakahashiAkamaHirokawa,TroelstraSchwichtenberg}.
In this strategy, the variable $\var{A}$
always refers to the closest binder above it. The 
substitution is the same
as that of the minimal strategy, but now the definition 
of the reduction
is modified in such a way
that $((\lambda_A t)~u)_{\Gamma}$ reduces to
$\underline{[u/A]}_{\varnothing}t$ instead of
$\underline{[u/A]}_{\Gamma}t$.
This way, the scheme $((\lambda_A \lambda_A \var{A})~t)_{\Gamma}$
reduces to $(\lambda_A \var{A})_{\Gamma}$, but the scheme
$(\lambda_A ((\lambda_A \var{A})~t))_{\Gamma}$ reduces to
$(\lambda_A t)_{\Gamma}$. 

The dual strategy is the {\em strategy with reference to the furthest
binder}. The substitution is the same as that of the maximal strategy,
but the definition of the reduction is modified in such a way that
$((\lambda_A t)~u)_{\Gamma}$ reduces to
$(\overline{[u/A]}_{\Gamma}t)_{\Gamma}$ when
$A \not\in \Gamma$ and to
$t_{\Gamma}$ when $A \in \Gamma$.
This way, the scheme $((\lambda_A \lambda_A \var{A})~t)_{\Gamma}$
reduces to $(\lambda_A t)_{\Gamma}$, but the scheme
$(\lambda_A ((\lambda_A \var{A})~t))_{\Gamma}$ reduces to
$(\lambda_A \var{A})_{\Gamma}$.

The counter-examples of Section 
\ref{counterex}
show that the maximal strategy and the strategy
with reference to the closest binder do not normalize, even if we
restrict to weak reduction, {\em i.e.} if we forbid reduction under
abstractions. We leave open the problem of the normalization of the
minimal strategy and the strategy with reference to the furthest
binder. However, we shall prove 
in Section \ref{secnorm}
the normalization of weak reduction 
for the minimal strategy.

Finally, $\lambda$-calculus is also a strategy of the scheme-calculus.
There, in order to reduce the scheme
$((\lambda_A \lambda_A \var{A})~t)_{\Gamma}$,
we need to know the history of the reduction, so that we are able to 
decide which 
binder the variable $\var{A}$ refers to.
In both schemes of type $A \Rightarrow A \Rightarrow A$ in the context $B$:
$$((\lambda_{A \Rightarrow A} \lambda_A \var{A \Rightarrow A})~(\lambda_A \var{A}))$$
and
$$\lambda_A ((\lambda_{B \Rightarrow A} \lambda_A (\var{B \Rightarrow A} \var{B}))~(\lambda_{B} \var{A}))$$
there is no ambiguity in the reference of the variable $\var{A}$ that appears
in the scope of a single binder of type $A$.

When we reduce these schemes, we get the normal scheme
$\lambda_A \lambda_A \var{A}$ in both cases.
But to determine the reference of the variable $\var{A}$ in this normal form,
we have to know where this
scheme is coming from. This is the role of variable names.
Calling $x$ the variable $\var{A}$, $y$ the variable $\var{B}$,
$f$ the variable $\var{A \Rightarrow A}$ and $g$ the variable
$\var{B \Rightarrow A}$, the first term
$((\lambda f \lambda x~f)~(\lambda x~x))$ reduces to
$\lambda x \lambda x~x$ and the second
$\lambda x~((\lambda g \lambda x~(g~y))~(\lambda y~x))$
to $\lambda x \lambda x'~x$, where a new name $x'$ has been introduced
by substitution to avoid the variable capture and keep the pointer from the
occurrence of the variable $x$ to its binder.

In this sense, the scheme-calculus generalizes both the lambda-calculus and the
total discharge calculus.

\section{A dependently-typed scheme-calculus}
\label{secdep}

The simply-typed scheme-calculus is much less expressive than
the simply-typed $\lambda$-calculus: with the general $\beta$-reduction the
uniqueness of normal forms is lost and if we restrict the calculus
to any deterministic strategy, as there is only one normal scheme of
type $A \Rightarrow A \Rightarrow A$, it is impossible to express
both projections.

In this section, we introduce a scheme-calculus with dependent types and 
prove that it is as expressive as the corresponding
typed $\lambda$-calculus. In particular, we construct a dependent
type system
that permits to express all the functions that are provably total in
impredicative arithmetic. This choice of impredicative arithmetic is
just an example and we could construct similar type systems for
various
theories.

\subsection{Terms and types}

We first define terms and types (or propositions) as
usual in many-sorted predicate logic.

We consider a language {\em i.e.} a set of sorts, a set
of function symbols each of them being equipped with an arity of the
form $\langle s_1, \ldots, s_n, s \rangle$, where $s_1, \ldots,
s_n, s$ are sorts, and a set of predicate symbols
each of them being equipped with an arity $\langle s_1, \ldots, s_n \rangle$, where
$s_1, \ldots, s_n$ are sorts. We consider also, for each sort, an infinite
set of variables.
The {\em terms of sort $s$} are inductively defined by
$$a = x~|~f(a_1,\ldots,a_n)$$
where $x$ is a variable of sort $s$, $f$ a function symbol of arity
$\langle s_1, \ldots, s_n, s \rangle$ and $a_1$, \ldots, $a_n$ are terms of
sorts $s_1, \ldots, s_n$, respectively.
The {\em types} are inductively defined by
$$A = P(a_1,\ldots,a_n)~|~A \Rightarrow A'~|~\fa x~A$$
where $P$ is a predicate symbol of arity $\langle s_1, \ldots, s_n \rangle$
and $a_1$, \ldots, $a_n$ are terms of sorts $s_1, \ldots, s_n$, respectively.

We could include other connectives and quantifiers and everything
would generalize smoothly. However, we prefer to define them in the
theory $HA_2$ presented in Section \ref{secHA2}.

Free and bound variables, alphabetic equivalence, as well as substitution are
defined as usual on terms and types. 

A {\em context} is a finite set of types.

To define a theory, such as arithmetic, we do not consider axioms.
Instead, we extend the natural deduction rules with a {\em conversion rule}
$$\irule{\Gamma \vdash A}
        {\Gamma \vdash B}
        {\mbox{$A \equiv B$ conv}}$$
allowing us to replace a proposition by an equivalent one for a
given congruence, at any time in a proof, like in {\em Deduction modulo}
\cite{DHK}.
The congruence has to be {\em non-confusing}, that is,
if $A \equiv B$ then either at least one of the propositions $A$, $B$ 
is atomic, or both are implications or both
are universal quantifications,
if $(A \Rightarrow A') \equiv (B \Rightarrow B')$ then
$A \equiv B$ and $A' \equiv B'$, and
if $(\fa x~A) \equiv (\fa x~B)$ then $A \equiv B$.

\subsection{Impredicative arithmetic}\label{secHA2}
\begin{figure}[t]
\noindent\framebox{\parbox{\columnwidth }{
$$x_1, \ldots, x_p~\epsilon_p~f_{\langle x_1, \ldots, x_p \rangle, \langle y_1, \ldots, y_n \rangle, A}(y_1, \dots, y_n) \lra A$$
$$x = y \lra \fa c~(x~\epsilon_1~c \Rightarrow y~\epsilon_1~c)$$
$$N(x) \lra \fa c~(0~\epsilon_1~c \Rightarrow 
\fa y~(N(y) \Rightarrow y~\epsilon_1~c \Rightarrow S(y)~\epsilon_1~c)
\Rightarrow x~\epsilon_1~c)$$
$$\nulll(0) \lra \fa c~(\epsilon_0(c) \Rightarrow \epsilon_0(c))$$
$$\nulll(S(x)) \lra \fa c~\epsilon_0(c)$$
$$\pred(0) \lra 0$$
$$\pred(S(x)) \lra x$$
\caption{The rewrite system $HA_2$\label{HA2}} }}
\end{figure}

Following \cite{DWPeano}, we can express predicative ({\em i.e.}
first-order) and impredicative ({\em i.e.} second-order) arithmetic in
Deduction modulo, hence the proofs of these theories can be expressed
in the scheme-calculus.

We introduce a sort $\iota$ for natural numbers
and a sort $\kappa_n$ ($n = 0, 1, 2, ...$) for $n$-ary classes of
natural numbers. The function symbols are $0$ (of sort $\iota$), $S$
and $\pred$ (of arity $\langle \iota, \iota \rangle$). The
predicate symbols are $=$ of arity $\langle \iota, \iota \rangle$,
$N$ and $\nulll$ of arity $\langle \iota  \rangle$ and $\epsilon_n$
of arity $\langle \iota, ..., \iota, \kappa_n \rangle$. We write
$p~\epsilon_1~c$ to express that the number $p$ is an element of the
(unary) class $c$, and $p_1, ..., p_n~\epsilon_n~c$ to express that the
sequence $p_1, ..., p_n$ is an element of the $n$-ary class $c$. 
Thus, $\epsilon_0(c)$ is the proposition corresponding to the
nullary class $c$. Moreover, for each proposition $A$, and sequences
of variables $\langle x_1, \ldots, x_p \rangle$, $\langle y_1,
\ldots, y_n \rangle$, such that the free variables of $A$ are among
$x_1, \ldots, x_p, y_1, \ldots, y_n$, we introduce a function symbol
$f_{\langle x_1, \ldots, x_p \rangle, \langle y_1, \ldots, y_n
\rangle, A}$ which is, informally speaking, obtained by Skolemizing
the instance of the comprehension scheme corresponding to $A$ with 
$x_1, ..., x_p$ as arguments of the class of arity $p$ and 
$y_1, ..., y_n$ as parameters.
Such
symbols exist for all propositions not containing Skolem symbols
themselves, in particular for propositions containing quantifiers on
classes (hence the impredicativity).

The meaning of these symbols is not expressed by axioms but by the 
rewrite rules in Figure \ref{HA2}. These rules define a congruence
on terms and propositions.

As is well known, the connectives and quantifiers
$\top$, $\bot$, $\neg$, $\wedge$, $\vee$, $\Leftrightarrow$, and $\ex$
can be defined in $HA_2$.
$$\begin{array}{rcl}
\top &=& \fa c~(\epsilon_0(c) \Rightarrow \epsilon_0(c))\\
\bot &=& \fa c~\epsilon_0(c)\\
\neg A  &=& A \Rightarrow \bot\\
A \wedge B &=& \fa c~((A \Rightarrow B \Rightarrow \epsilon_0(c))
\Rightarrow \epsilon_0(c))\\
A \vee B &=& \fa c~((A \Rightarrow \epsilon_0(c))
\Rightarrow (B \Rightarrow \epsilon_0(c))
\Rightarrow \epsilon_0(c))\\
A \Leftrightarrow B &=& (A \Rightarrow B) \wedge (B \Rightarrow A)\\
\ex x~A &=& \fa c~((\fa x~(A \Rightarrow \epsilon_0(c)))
\Rightarrow \epsilon_0(c))
\end{array}$$

Using the congruence defined by the rules in Figure \ref{HA2}
and the conversion rule, the usual axioms of
impredicative arithmetic, can easily be proven.
$$\fa x~(x = x)$$
$$\fa x \fa y \fa c~(x = y \Rightarrow x~\epsilon_1~c \Rightarrow
y~\epsilon_1~c)$$
$$N(0)$$
$$\fa x~(N(x) \Rightarrow N(S(x)))$$
$$\fa x \fa y~(S(x) = S(y) \Rightarrow x = y)$$
$$\fa x~\neg (0 = S(x))$$
$$\fa c~(0~\epsilon_1~c \Rightarrow
\fa y~(N(y) \Rightarrow y~\epsilon_1~c \Rightarrow
S(y)~\epsilon_1~c) \Rightarrow \fa x~(N(x) \Rightarrow x~\epsilon_1~c)$$
$$\fa y_1 ... \fa y_n \ex c \fa x_1 ... \fa x_p((x_1 ... x_p~\epsilon_p~c) \Leftrightarrow A)$$
where $A$ is any proposition
not containing Skolem symbols, and whose free variables are among
$y_1, ... ,y_n, x_1, ..., x_p$.

\subsection{Schemes}

\begin{definition}[Schemes]
Schemes are inductively defined as follows.
$$t = \var{A}~|~\lambda_{A} t~|~(t~t')~|~\Lambda x~t~|~(t~a)$$
\end{definition}
Each construct corresponds to a natural deduction rule.
Typing rules are given in Figure \ref{typing}.
They are the rules of natural deduction.

\begin{figure}[t]
\noindent\framebox{\parbox{\columnwidth }{
$$\irule{}
        {\Gamma \vdash \var{A} :A}
        {\mbox{$A \in \Gamma$~ax}}$$
$$\irule{\Gamma \cup \{A\} \vdash t: B}
        {\Gamma \vdash \lambda_{A} t : A \Rightarrow B}
        {\mbox{$\Rightarrow$i}}$$
$$\irule{\Gamma \vdash t:  A \Rightarrow B~~~\Gamma \vdash u: A}
        {\Gamma \vdash (t~u):B}
        {\mbox{$\Rightarrow$e}}$$
$$\irule{\Gamma \vdash t: A}
        {\Gamma \vdash \Lambda x~t: \fa x~A}
        {\mbox{$x \not\in FV(\Gamma)$ $\fa$i}} $$
$$\irule{\Gamma \vdash t: \fa x~A}
        {\Gamma \vdash (t~a) : [a/x]A}
        {\mbox{$\fa$e}}$$
$$\irule{\Gamma \vdash t: A}
        {\Gamma \vdash t: B}
        {\mbox{$A \equiv B$~conv}}$$
\caption{Dependently-typed schemes \label{typing}} }}
\end{figure}

\begin{definition}[Scheme in context]
A {\em scheme in context} is a pair $t_{\Gamma}$ where $t$ is a
scheme and $\Gamma$ is a context such that $t$ is well-typed in 
$\Gamma$.
\end{definition}
We sometimes omit the context $\Gamma$ when there is no ambiguity.

\medskip

We now define the reduction relation on schemes. Before that, we
define the application of a substitution of term variables and that of
scheme variables to a scheme.

\begin{definition}
Let $\theta$ be a substitution of term variables and $t$ be a
scheme. The scheme $\theta t$ is inductively defined by
\begin{itemize}
\item $\theta \var{A} = \var{\theta A}$,
\item $\theta (\lambda_A t) = \lambda_{\theta A} \theta t$,
\item $\theta (u~v) = (\theta u~\theta v)$,
\item $\theta (\Lambda x~t) = \Lambda x'~(\theta [x'/x]t)$,
where $x'$ is a variable which occurs neither in $\Lambda x~t$ nor
in $\theta$,
\item $\theta (t~a) = (\theta t~\theta a)$.
\end{itemize}
\end{definition}

Remark that this substitution, as usual, avoids variable capture by
renaming bound term variables.

A substitution of scheme variables is a function of finite domain
associating schemes to types. The application of a
substitution to a scheme with respect to a context is defined as follows.

\begin{definition}[Substitution]
\label{defsubstdep}
~
\begin{itemize}
\item $\sigma_{\Gamma} \var{A} = \{\var{A}, \sigma(A)\}$,
if $A \in dom(\sigma)$ and $A \in \Gamma$,
\item $\sigma_{\Gamma} \var{A} = \{\sigma(A)\}$,
if $A \in dom(\sigma)$ and $A \not\in \Gamma$,
\item $\sigma_{\Gamma} \var{A} = \{\var{A}\}$,
if $A \not\in dom(\sigma)$,
\item $\sigma_{\Gamma} \lambda_A t = \lambda_A \sigma_{(\Gamma \cup \{A\})} t$,
\item $\sigma_{\Gamma} (u~v) = (\sigma_{\Gamma} u~\sigma_{\Gamma} v)$,
\item $\sigma_{\Gamma} \Lambda x~t = \Lambda x'~\sigma_{\Gamma} [x'/x]t$,
where $x'$ is a variable that occurs neither in $\Lambda x~t$ nor in
$\sigma$,
\item $\sigma_{\Gamma} (t~a) = (\sigma_{\Gamma} t~a)$.
\end{itemize}
\end{definition}

\begin{definition}[Reduction]
\label{defreddep}
The {\em one step top level $\beta$-reduction} is defined by the rules

\begin{itemize}
\item $((\lambda_A t)~u)_{\Gamma} \lra v_{\Gamma}$,
for all $v \in [u/A]_{\Gamma} t$,
\item $((\Lambda x~t)~a)_{\Gamma} \lra ([a/x]t)_{\Gamma}$.
\end{itemize}
The {\em one step $\beta$-reduction} relation $\rhd$ is
the contextual closure of this relation and the {\em
$\beta$-reduction} relation $\rhd^*$ is the
reflexive-transitive closure of the relation $\rhd$.
\end{definition}

\section{Subject-reduction}
\label{secsubred}

\begin{proposition}[Substitution]
\label{lemmsubstitution}
~
\begin{itemize}
\item If $\Gamma \vdash t : B$, then $[a/x]\Gamma \vdash [a/x]t :
[a/x]B$.
\item If $\Gamma \cup \{A\} \vdash t : B$ and $\Gamma \vdash u : A$, then $\Gamma \vdash v : B$, for
all $v$ in $[u/A]_{\Gamma} t$.
\end{itemize}
\end{proposition}

\proof{By induction over the structure of $t$.}

\medskip

Remark that this substitution lemma holds although
bound variables may be substituted and variable
capture is allowed. That is because the captured variables have the
same type as the binder that captures them.

\begin{proposition}[Inversion]
\label{inversion} Let $\Gamma \vdash t : A$.
\begin{enumerate}
\item
If $t$ is some variable $\var{B}$, then $\Gamma$ contains the
proposition $B$ and $A \equiv B$,

\item
If $t = \lambda_B u$, then there exists a type $C$ such that $\Gamma \cup \{B\} 
\vdash u : C$ and $A \equiv (B \Rightarrow C)$.

\item
If $t = (u~v)$, where $u$ and $v$ are schemes, then there exist types
$B$ and $C$ such that $\Gamma \vdash u : B \Rightarrow C$ and $\Gamma
\vdash v : B$, and $A \equiv C$ .

\item
If $t = \Lambda x~u$, then there exists a variable $x$ and a type $B$ 
such that $\Gamma \vdash u : B$ and 
$A \equiv (\fa x~B)$ and $x \not\in FV(\Gamma)$.

\item
If $t = (u~a)$, where $u$ is a scheme and $a$ a term, then there
exists a type $B$ such that $\Gamma \vdash u : \fa x~B$ and $A \equiv
[a/x]B$.
\end{enumerate}
\end{proposition}

\proof{By induction on the typing derivation. If the last rule
is conversion, we apply the induction hypothesis and
the transitivity of $\equiv$. Otherwise the
premises of the rule yield the result.}

\medskip

We are now ready to prove the subject-reduction property. Before that, 
we need to prove the proposition below.

\begin{proposition}
\label{propsubjectreduction}
If $\Gamma \vdash t:A$ and $t_{\Gamma} \lra u_{\Gamma}$, then $\Gamma \vdash u:A$.
\end{proposition}

\proof{If $t = ((\lambda_B t_1)~t_2)$ and $u \in [t_2/B]_{\Gamma}t_1$, then by
Proposition \ref{inversion}({\em 3}), there exist types $B'$ and $C'$ such
that $\Gamma \vdash \lambda_B t_1:B' \Rightarrow C'$, $\Gamma
\vdash t_2:B'$ and $A \equiv C'$ and by Proposition \ref{inversion}({\em 2}),
there exists a type $C$ such that $\Gamma \cup \{B\} \vdash t_1: C$ and $B'
\Rightarrow C' \equiv B \Rightarrow C$.
As the congruence is non-confusing, we have $B \equiv B'$ and $C \equiv C'$.
Using the conversion rule, we have $\Gamma \vdash t_2:B$. By Proposition
\ref{lemmsubstitution}, we get $\Gamma \vdash u: C$ and using the conversion
rule, $\Gamma \vdash u: A$.

If $t = ((\Lambda x~t_1)~a)$ and $u = [a/x]t_1$, we
choose $x$ not occurring in $\Gamma$.
Using Proposition \ref{inversion}, non-confusion and
conversion, we get a type $B'$ such that
$\Gamma \vdash t_1:B'$ and $A \equiv [a/x]B'$.
We conclude with Proposition \ref{lemmsubstitution} and conversion.}

\begin{theorem}[Subject-reduction]
If $\Gamma \vdash t:A$ and $t_{\Gamma} \rhd^* u_{\Gamma}$, then
$\Gamma \vdash u:A$.
\end{theorem}

\proof{We show, by induction on the derivation of $t_{\Gamma} \rhd u_{\Gamma}$,
that if $\Gamma \vdash t:A$ and $t_{\Gamma} \rhd u_{\Gamma}$
then $\Gamma \vdash u:A$ and we conclude by induction on the
length of reduction sequences.}

\section{Weak normalization of weak reduction}
\label{secnorm}

We now prove that each scheme can be reduced to a normal
form. Because of the counter-examples given in Section
\ref{secsimple}, we cannot expect to prove strong normalization for
the reduction of the scheme-calculus. Of course, it is possible to
prove weak normalization by mimicking the reductions of
$\lambda$-calculus. But the scheme reduction strategy provided by the
proof of the normalization theorem is as important as the theorem itself and
the strategy provided by this trivial proof would require to introduce
variable names, which is precisely what we want to
avoid. Thus, we shall give another normalization proof which provides
a strategy that can be defined without introducing variable names.

The first step towards a normalization result is to restrict
substitution to minimal substitution, {\em i.e.} to modify the
first clause of Definition \ref{defsubstdep}: instead of taking the
clause
\begin{itemize}
\item $\sigma_{\Gamma} \var{A} = \{\var{A}, \sigma(A)\}$,
if $A \in dom(\sigma)$ and $A \in \Gamma$
\end{itemize}
we take the following one
\begin{itemize}
\item $\underline{\sigma}_{\Gamma} \var{A} = \{\var{A}\}$,
if $A \in dom(\sigma)$ and $A \in \Gamma$.
\end{itemize}

Restricting substitution to minimal substitution rules out the
counter-examples of Section \ref{secsimple}. Moreover, minimal
substitution enjoys several properties of substitution of
$\lambda$-calculus. In particular, bound variables are never
substituted. Thus, we conjecture this minimal reduction to be
strongly normalizing. However, we shall leave this problem open and
prove a slightly weaker result: the normalization of weak reduction,
{\em i.e.} of the reduction where reduction is not performed under
abstractions. Indeed, the minimal reduction lacks one property of the
reduction of
$\lambda$-calculus: the commutation of reduction and substitution, {\em i.e.}
that whenever $t_{\Gamma \cup \{A\}}$ 
reduces to $u_{\Gamma \cup \{A\}}$ and $v$ is a scheme of type $A$ 
in the context 
$\Gamma$, then
$(\underline{[v/A]}_{\Gamma}t)_{\Gamma}$ reduces to
$(\underline{[v/A]}_{\Gamma}u)_{\Gamma}$.
For instance, if $\Gamma = \{A \Rightarrow A, B \Rightarrow A, B\}$, 
$t = ((\lambda_A \var{A})~(\var{A \Rightarrow A}~\var{A}))$, 
$u = \var{A}$ and $v = (\var{B \Rightarrow A}~\var{B})$, 
then $t_{\Gamma \cup   \{A\}}$ reduces to $u_{\Gamma \cup \{A\}}$. 
But
$(\underline{[v/A]}_{\Gamma}t)_{\Gamma} 
= ((\lambda_A \var{A})~(\var{A \Rightarrow A}~(\var{B \Rightarrow A}~\var{B})))_{\Gamma}$
reduces to
$(\var{A \Rightarrow A}~(\var{B \Rightarrow A}~\var{B}))$
and not to
$(\underline{[v/A]}_{\Gamma}u)_{\Gamma}
= 
(\var{B \Rightarrow A}~\var{B})$.

This property is
unfortunately needed in normalization proofs for strong reduction
based on reducibility candidates. But, it is not needed, if we
restrict to weak reduction.

On the other hand, the normalization of weak reduction is sufficient
to prove the existence of weak head normal forms, which is itself
sufficient to extract witnesses from existential proofs.

The proof presented in this section is based on ideas similar to
those of \cite{DW}. The main difference is that we take into account that
reduction does not commute with substitution.

\subsection{Reduction}

The {\em one step minimal top level reduction} 
$\lra_{\mbox{\it min}}$
is defined as
in Definition \ref{defreddep} except that substitution is minimal
substitution. Instead of considering the contextual closure of this relation, 
we define the {\em one step weak minimal reduction} as follows.

\begin{definition}[Weak minimal reduction]
The {\em one step weak minimal reduction} 
$\twoheadrightarrow$ is defined by considering 
any abstraction and any application whose left-hand side is normal, 
as a normal form, otherwise by reducing the leftmost 
reduct. 
It is inductively
defined as follows. 

Let $t$ and $u$ be schemes and $a$ be a term, 
\begin{itemize}
\item if $t_{\Gamma} \lra_{\mbox{min}} u_{\Gamma}$ then 
$t_{\Gamma} \twoheadrightarrow u_{\Gamma}$,

\item if $t_{\Gamma} \twoheadrightarrow t'_{\Gamma}$, then $(t~u)_{\Gamma} \twoheadrightarrow (t'~u)_{\Gamma}$,

\item if $t_{\Gamma} \twoheadrightarrow t'_{\Gamma}$, then $(t~a)_{\Gamma} \twoheadrightarrow (t'~a)_{\Gamma}$.
\end{itemize}

The {\em weak minimal reduction} relation $\twoheadrightarrow^*$ is the
reflexive-transitive closure of the relation $\twoheadrightarrow$.
\end{definition}

Notice that the  relation $\twoheadrightarrow$
is functional ({\em i.e.} deterministic) in the sense that,
for each scheme $t$, there is at most one scheme $t'$ such that 
$t \twoheadrightarrow t'$.

The reduction sequence issued from $t_{\Gamma}$ is the (finite or infinite)
sequence $t_{0, \Gamma}, t_{1, \Gamma}, t_{2, \Gamma}, \ldots$ such that
$t_{0, \Gamma} = t_{\Gamma}$,
and for all $i$, if there exists a $t'$ such that
$t_{i, \Gamma} \twoheadrightarrow t'_{\Gamma}$,
then the sequence is defined at $i+1$ and $t_{i+1, \Gamma} = t'_{\Gamma}$,
otherwise $t_{i, \Gamma}$ is the last element of the sequence.
A scheme in context $t_{\Gamma}$ is said to be {\em normalizing} if
its reduction sequence is finite. Hereafter, we write ${\cal N}$ for the set 
of normalizing schemes in contexts.

\begin{proposition}[Properties of minimal substitution]
 \label{propminimalsubst}
~
\begin{enumerate}
\item If $t$ is well-typed in $\Gamma$, then
$\underline{[w/A]}_{\Gamma} t = t$.
\item If $A \in \Gamma$, then $\underline{[w/A]}_{\Gamma} t = t$.
\item If $B \neq A$, then $\underline{[w/A]}_{\Gamma \cup \{B\}} t =
\underline{[w/A]}_{\Gamma} t$.
\end{enumerate}
\end{proposition}

\proof{
\begin{enumerate}
\item
By induction on the structure of $t$. The only non-trivial case
is when $t = \var{B}$. In this case, $B \in \Gamma$ and
both schemes are equal to $\var{B}$.

\item
By induction on the structure of $t$. The only non-trivial case is
when $t = \var{A}$. In this case $A \in \Gamma$ and
thus both schemes are equal to $\var{A}$.

\item
By induction on the structure of $t$. The only non-trivial case
is when $t = \var{A}$. In this case, either $A \in \Gamma$
in which case both schemes are equal to $\var{A}$ or
$A \not\in \Gamma$, in which case $A \not\in (\Gamma \cup \{B\})$
and both schemes are equal to $w$.
\end{enumerate}}

\subsection{Girard's reducibility candidates}

\begin{definition}[Operations on sets of schemes]

If $E$ and $F$ are sets of schemes in contexts, we define
the set
$$E~\tilde{\Rightarrow}~F = \{t_{\Gamma} \in {\cal N}~|~\fa
t' \fa u~((t_{\Gamma} \twoheadrightarrow^* (\lambda_A
t')_{\Gamma}~\mbox{and}~u_{\Gamma} \in E)
\Rightarrow (\underline{[u/A]}_{\Gamma} t')_{\Gamma} \in F)\}$$
If $S$ is a set of sets of schemes in contexts, we define the set
$$\tilde{\fa}~S =
\{t_{\Gamma} \in {\cal N}~|~\fa t' \fa a \fa E~((t_{\Gamma}
\twoheadrightarrow^* (\Lambda x~t')_{\Gamma}~\mbox{and}~E \in S) \Rightarrow ([a/x]t')_{\Gamma} \in E)\}$$
\end{definition}

\begin{definition}[Reducibility candidate \cite{Girard}]
\label{defCR}
A scheme is said to be {\em neutral} if it corresponds to an axiom rule
or an elimination rule,
but not to an introduction rule. A set $R$ of schemes in contexts
is said to be a {\em reducibility candidate}, if the following conditions are satisfied:
\begin{itemize}
\item if $t_{\Gamma} \in R$, then $t_{\Gamma}$ is normalizing,
\item if $t_{\Gamma} \in R$ and $t_{\Gamma} \twoheadrightarrow^* t'_{\Gamma}$,
then $t'_{\Gamma} \in R$,
\item if $t_{\Gamma}$ is neutral,
and for every $t'_{\Gamma}$ such that $t_{\Gamma} \twoheadrightarrow t'_{\Gamma}$, we
have $t'_{\Gamma} \in R$, then $t_{\Gamma} \in R$.
\end{itemize}
We write ${\cal C}$ for the set of reducibility candidates.
\end{definition}

Remark that, as the reduction relation is deterministic, the third condition
can be rephrased as:
(1) if $t_{\Gamma}$ is neutral and normal, then $t_{\Gamma} \in R$,
and
(2) if $t_{\Gamma}$ is neutral, has a one-step reduct $t'_{\Gamma}$
and this reduct is in $R$, then $t_{\Gamma}$ is in $R$.

\begin{proposition}
If $E$ and $F$ are sets of schemes in contexts, then
$E~\tilde{\Rightarrow}~F$ is a
reducibility candidate.
If $S$ is a set of sets of schemes in contexts, then
$\tilde{\fa}~S$ is a reducibility candidate.
\end{proposition}

\proof{By definition, all the schemes in the sets
$E~\tilde{\Rightarrow}~F$ and $\tilde{\fa}~S$ are normalizing.

For closure by reduction, just remark that if $t_{\Gamma} \twoheadrightarrow^*
t'_{\Gamma}$ and $t_{\Gamma}$ is normalizing, then so is
$t'_{\Gamma}$ and that if $t_{\Gamma} \twoheadrightarrow^* t'_{\Gamma}$ and
$t'_{\Gamma} \twoheadrightarrow^* u_{\Gamma}$, then
$t \twoheadrightarrow^* u_{\Gamma}$.

For the third property, remark that if $t_{\Gamma}$ is a scheme
in context and for all $t'_{\Gamma}$ such that $t_{\Gamma} \twoheadrightarrow t'_{\Gamma}$,
$t'_{\Gamma}$ is
normalizing then $t_{\Gamma}$ is normalizing and that if $t_{\Gamma}$ is a
neutral scheme in context and $t_{\Gamma} \twoheadrightarrow^* u_{\Gamma}$ where $u$ is an
introduction, then the reduction sequence is not empty, thus there
exists a scheme $t'_{\Gamma}$ such that $t_{\Gamma} \twoheadrightarrow t'_{\Gamma} \twoheadrightarrow^*
u_{\Gamma}$.}

\subsection{${\cal C}$-models}

A {\em model valued in the algebra of reducibility candidates}, 
or {\em ${\cal C}$-model}, is defined as a classical model except 
that propositions are interpreted in the algebra ${\cal C}$ of
reducibility candidates. Thus it consists of 
a set $M_s$, for each sort $s$, a function
$\hat{f}$ from $M_{s_1} \times \dots \times M_{s_n}$ to $M_{s}$,
for each function symbol $f$ of arity $\langle s_1,\dots,s_n,s \rangle$,
and a function $\hat{P}$ from 
$M_{s_1} \times \dots \times M_{s_n}$ to ${\cal C}$,
for each predicate symbol $P$ of arity $\langle s_1,\dots,s_n \rangle$. 
The denotation of terms in a valuation is defined as usual. The denotation 
of propositions is defined by 
\begin{itemize}
\item $\llbracket P(a_1, \ldots, a_n) \rrbracket_{\phi}
= \hat{P}(\llbracket a_1 \rrbracket_{\phi}, \ldots,
\llbracket a_n \rrbracket_{\phi})$,

\item $\llbracket A \Rightarrow B \rrbracket_{\phi} =
\llbracket A \rrbracket_{\phi}~\tilde{\Rightarrow}~
\llbracket B \rrbracket_{\phi}$,

\item
$\llbracket \fa x~A \rrbracket_{\phi} =
\tilde{\fa}~\{\llbracket A \rrbracket_{\phi + x = e}~|~e \in M_s\}$,
where $s$ is the sort of the variable $x$ and $\phi + x = e$ is 
the valuation coinciding with $\phi$ everywhere except in $x$ where it
takes the value $e$.
\end{itemize}

\begin{definition}
A congruence $\equiv$ is said to be valid in a ${\cal C}$-model
${\cal M}$ if for all types $A$ and $B$, and every
valuation $\phi$, $A \equiv B$ implies $\llbracket A
 \rrbracket_{\phi} = \llbracket B \rrbracket_{\phi}$.
\end{definition}

\subsection{Weak normalization of weak reduction}

As variable captures are allowed in the scheme calculus, the substitutions 
cannot be composed as usual. For instance if $\Gamma = \{B \Rightarrow A, 
C \Rightarrow B, C\}$, $f$ is the variable 
$\var{B \Rightarrow A}$ and $u$ is the term $(\var{C \Rightarrow B}~\var{C})$, 
we have 
$$[u/B]_{\Gamma}[(f~\var{B})/A]_{\Gamma \cup \{B\}}
(\lambda_B~\var{A}) = (\lambda_B~(f~\var{B}))$$
and 
$$[u/B]_{\Gamma}[(f~\var{B})/A]_{\Gamma \cup \{B\}}\var{A} = (f~u)$$
but there is no substitution $\sigma$ and context $\Delta$ such that 
$$\sigma_{\Delta}(\lambda_B~\var{A}) = (\lambda_B~(f~\var{B}))$$
and 
$$\sigma_{\Delta} \var{A} = (f~u)$$
because we cannot have at the same time 
$\sigma_{\Delta \cup \{B\}} \var{A} = (f~\var{B})$ 
and
$\sigma_{\Delta} \var{A} = (f~u)$.
Thus arises the need for the notion of {\em free sequence of substitutions}. 

\begin{definition}[Free sequence of substitutions]
Let $\Gamma$ be a context, and $\phi$ be a valuation, the free
sequences of substitutions in $\Gamma, \phi$ are inductively defined
as follows.
\begin{itemize}
\item The empty sequence is a free sequence of substitutions.
\item If $\rho$ is a free sequence of substitutions,
$C$ is a type,
$w$ is a scheme in the context $\Gamma$, such that
$w_{\Gamma} \in \llbracket C \rrbracket_{\phi}$,
then $(\underline{[w/C]}_{\Gamma}, \rho)$ is a free sequence of substitutions.
\item If $\rho$ is a free sequence of substitutions, $x$ is a term variable
that does not occur in $\rho$, and $a$ is a term, then
$([a/x], \rho)$ is a free sequence of substitutions.
\end{itemize}
\end{definition}

\begin{definition}
Let $\rho$ be a free sequence of substitutions in $\Gamma, \phi$ and
$a$ be a term. 

The term $\rho a$ is defined as follows.
\begin{itemize}
\item If $\rho$ is the empty sequence, then $\rho a = a$,
\item If $\rho = (\underline{[w/C]}_{\Gamma}, \rho')$,
then $\rho a = \rho' a$,
\item If $\rho = ([b/x], \rho')$, then $\rho a = [b/x](\rho' a)$.
\end{itemize}

Let $A$ be a type,
the type $\rho A$ is defined as follows.
\begin{itemize}
\item If $\rho$ is the empty sequence, then $\rho A = A$,
\item If $\rho = (\underline{[w/C]}_{\Gamma}, \rho')$,
then $\rho A = \rho' A$,
\item If $\rho = ([b/x], \rho')$, then $\rho A = [b/x](\rho' A)$.
\end{itemize}

Let $t$ be a scheme,
the scheme $\rho t$ is defined as follows.
\begin{itemize}
\item If $\rho$ is the empty sequence, then $\rho t = t$,
\item If $\rho = (\underline{[w/C]}_{\Gamma}, \rho')$,
then $\rho t = \underline{[w/\rho' C]}_{\Gamma}(\rho' t)$,
\item If $\rho = ([b/x], \rho')$, then $\rho t = [b/x](\rho' t)$.
\end{itemize}
\end{definition}

In the proposition below, we prove, as usual, that if a scheme has type $A$ then 
it is an element of the interpretation of $A$ (hence we shall be able to 
deduce that it is normalizing). 

\begin{proposition}
\label{main}
Let $\equiv$ be a congruence,
${\cal M}$ be a ${\cal C}$-model of $\equiv$,
$\Gamma$ and $\Delta$ be contexts,
$\phi$ be a valuation,
$t$ be a scheme of type $A$ modulo $\equiv$ in $\Delta$,
and $\rho$ be a free sequence of substitutions in $\Gamma, \phi$
such that $\rho t$ is a scheme well-typed in $\Gamma$.
Then $(\rho t)_{\Gamma} \in \llbracket A \rrbracket_{\phi}$.
\end{proposition}

\proof{
By induction on the typing derivation of $t$.
\begin{itemize}

\item {\em ax.} The scheme $t$ is equal to $\var{A}$. If $A$ is
not in the domain of any substitution of $\rho$ or $A \in \Gamma$, then
$(\rho t)_{\Gamma} = \var{\rho A}_{\Gamma}$. Thus, as the candidate
$\llbracket A \rrbracket_{\phi}$ contains all normal neutral schemes,
$(\rho t)_{\Gamma} \in \llbracket A \rrbracket_{\phi}$.
Otherwise, let $[w/A]_{\Gamma}$ be the
rightmost substitution of $\rho$ binding $A$. We have
$\rho = \rho_2, [w/A]_{\Gamma}, \rho_1$ and
$(([w/A]_{\Gamma}, \rho_1) \var{A})_{\Gamma}
= ([w/\rho_1A]_{\Gamma}, \var {\rho_1 A})_{\Gamma} = w_{\Gamma}$.
The sequence $\rho$ is a free sequence of substitutions,
the scheme $w$ is well-typed in $\Gamma$, and it
does not contain any term variable bound in $\rho_2$,
thus, using Proposition \ref{propminimalsubst}({\em 1}), $(\rho t)_{\Gamma}
= w_{\Gamma} \in \llbracket A
\rrbracket_{\phi}$.

\item {\em $\Rightarrow$i.}
The scheme $t$ has the form $\lambda_B u$,
$A = (B \Rightarrow B')$ and
$(\rho t)_{\Gamma} = (\rho (\lambda_B u))_{\Gamma}$.
Traversing the abstraction, the substitutions of $\rho$ have their context
extended to $\Gamma \cup \{C\}$ for some type $C$.
Using Proposition \ref{propminimalsubst}({\em 2}),
we drop those substitutions in $\rho$
that bind the type $C$ and using Proposition
\ref{propminimalsubst}({\em 3}),
we erase $C$ from the context of the remaining ones.
We get this way another
free sequence of substitutions $\rho'$ in $\Gamma, \phi$ and
$(\rho t)_{\Gamma} = (\lambda_{\rho B} (\rho' u))_{\Gamma}
= (\lambda_{\rho' B} (\rho' u))_{\Gamma}$.
This scheme is normal, hence it is normalizing and it only
reduces to itself.
To prove that it is in $\llbracket A \rrbracket_{\phi} =
\llbracket B \Rightarrow B' \rrbracket_{\phi}$,
we need to prove that
for all schemes $v$ in $\Gamma$ such that
$v_{\Gamma} \in \llbracket B \rrbracket_{\phi}$, the scheme
$([v/\rho' B]_{\Gamma} (\rho' u))_{\Gamma} = (([v/B]_{\Gamma},\rho') u)_{\Gamma}$
is in $\llbracket B' \rrbracket_{\phi}$. This follows
from induction hypothesis and the fact that
$([v/B]_{\Gamma}, \rho')$ is a free sequence of substitutions.

\item {\em $\fa$i.}
The scheme $t$ has the form $\Lambda x~u$,
we can assume, without loss of generality, that
$x$ does not occur in $\rho$. We have $A = (\fa x~B)$
and $\rho t = \Lambda x~\rho u$. This scheme
is normal, hence it is normalizing and it only
reduces to itself.
To prove that it is in $\llbracket A \rrbracket_{\phi} =
\llbracket \fa x~B \rrbracket_{\phi}$,
we need to prove that
for all terms $a$, and $e$ in $M_s$, where $s$ is the sort of the
variable $x$, the scheme
$([a/x] (\rho u))_{\Gamma} = (([a/x],\rho) u)_{\Gamma}$ is in
$\llbracket B \rrbracket_{\phi + x = e}$.
As $x$ does not occur in $\rho$, the sequence
$([a/x], \rho)$ is a free sequence of substitutions for $\Gamma, (\phi + x = e)$.
Thus, this scheme is in $\llbracket B \rrbracket_{\phi + x = e}$
by induction hypothesis.

\item {\em $\Rightarrow$e.}
The scheme $t$ has the form $(u~v)$. Thus,
$\rho t = (u'~\rho v)$,
where $u' = \rho u$.
By induction hypothesis,
$u'_{\Gamma} \in \llbracket B \Rightarrow A \rrbracket_{\phi}$
and $(\rho v)_{\Gamma} \in \llbracket B \rrbracket_{\phi}$.
Thus, the scheme $u'_{\Gamma}$ is normalizing. Let $n$ be the length
of the reduction sequence starting from $u'_{\Gamma}$.
We prove, by induction on $n$ that if
$u'_{\Gamma} \in \llbracket B \Rightarrow A \rrbracket_{\phi}$ and
the length of the reduction sequence starting from $u$ is $n$, and
$v'_{\Gamma} \in \llbracket B \rrbracket_{\phi}$ then $(u'~v')_{\Gamma}
\in \llbracket A \rrbracket_{\phi}$.
As $(u'~v')_{\Gamma}$ is neutral, all we
need to prove is that its potential one-step reduct
is in $\llbracket A \rrbracket_{\phi}$. If the reduction takes place in $u'$,
we just apply the induction hypothesis. Otherwise, the reduction takes place
at top level. We have $u'_{\Gamma} = (\lambda_{\rho B} u'')_{\Gamma}$
and the reduct is
$([v'/\rho B]_{\Gamma} u'')_{\Gamma}$
which is in $\llbracket A \rrbracket_{\phi}$ by definition of
$\llbracket B \Rightarrow A \rrbracket_{\phi}$.

\item {\em $\fa$e.}
The scheme $t$ has the form $(u~a)$, where $u$ has type $\fa x~B$,
$A = [a/x]B$, $\rho t = (u'~\rho a)$, where $u' = \rho u$.
By induction hypothesis,
$u'_{\Gamma} \in \llbracket \fa x~B \rrbracket_{\phi}$.
Thus, the scheme $u'_{\Gamma}$ is normalizing. Let $n$ be the length
of the reduction sequence starting from this scheme.
We prove, by induction on $n$ that if
$u'_{\Gamma} \in \llbracket \fa x~B \rrbracket_{\phi}$,
the length of the reduction sequence starting from $u$ is $n$,
and $a'$ is a term,
then $(u'~a')_{\Gamma}
\in \llbracket [a/x]B \rrbracket_{\phi} = \llbracket B
\rrbracket_{\phi + x = \llbracket a \rrbracket_{\phi}}$.
As this scheme is neutral, all we
need to prove is that its potential one-step reduct
is in $\llbracket B \rrbracket_{\phi + x = \llbracket a \rrbracket_{\phi}}$.
If the reduction takes place in $u'$,
we just apply the induction hypothesis. Otherwise, the reduction takes place
at top level. We have $u'_{\Gamma} = (\Lambda x~u'')_{\Gamma}$ and the reduct is
$([a'/x]u'')_{\Gamma}$
which,
by definition of
$\llbracket \fa x~B \rrbracket_{\phi}$,
is in $\llbracket B \rrbracket_{\phi + x = \llbracket a \rrbracket_{\phi}}$.

\item {\em conv.} If the last rule is a conversion rule,
by induction hypothesis, we have
$(\rho t)_{\Gamma} \in \llbracket B \rrbracket_{\phi}$ for some
$B \equiv A$, and we have
$\llbracket B \rrbracket_{\phi} = \llbracket A \rrbracket_{\phi}$.
Thus $(\rho t)_{\Gamma} \in \llbracket A \rrbracket_{\phi}$.
\end{itemize}}

\begin{theorem}[Normalization]
\label{thnorm}
Let $\equiv$ be a congruence that has a ${\cal C}$-model ${\cal M}$.
Let $\Gamma$ be a context and $t$ a scheme of type $A$ modulo $\equiv$
in $\Gamma$. Then $t_{\Gamma}$ is normalizing.
\end{theorem}

\proof{By Proposition \ref{main}, for all $\phi$,
$t_{\Gamma} \in \llbracket A \rrbracket_{\phi}$, thus it is normalizing.}

\subsection{Normalization in $HA_2$}

\begin{proposition}
\label{normHA2}
All schemes well-typed in $HA_2$ are normalizing.
\end{proposition}

\proof{We construct a ${\cal C}$-model as follows.
Let $M_{\iota} =
{\mathbb N}$ and $M_{\kappa_n} = {\mathbb N}^n \rightarrow {\cal
C}$.
The symbols $0$, $S$, and $\pred$ are
interpreted in the standard way.
The function
$\hat{\epsilon}_n$ maps $k_1, \ldots, k_n$ and $f$ to $f(k_1, \ldots,
k_n)$,
$\hat{=}$ maps $n$ and $m$ to
$\llbracket \fa c~(x~\epsilon_1~c \Rightarrow y~\epsilon_1~c)
 \rrbracket_{n/x,m/y}$ and $\hat{\nulll}$ maps $0$ to
$\llbracket \fa c~(\epsilon_0(c) \Rightarrow \epsilon_0(c)) \rrbracket$
and the other numbers to $\llbracket \fa c~\epsilon_0(c) \rrbracket$.

To define $\hat{N}$, we first define the function
$\Phi$ that maps any
function $\alpha$ of ${\mathbb N} \rightarrow {\cal C}$ to the
function that maps $n$ to
the interpretation of the proposition
$\fa c~(0~\epsilon_1~c \Rightarrow \fa y~(N(y) \Rightarrow
y~\epsilon_1~c \Rightarrow S(y)~\epsilon_1~c) \Rightarrow
x~\epsilon_1~c)$, for the valuation $n/x$,
in the model of domains $M_{\iota}$ and
$M_{\kappa_n}$, and where $0$ and $S$ are interpreted in the standard
way, $\epsilon_n$ is interpreted by $\hat{\epsilon}_n$,
but $N$ is interpreted by $\alpha$.
The set ${\mathbb N} \ra {\cal C}$ ordered by pointwise inclusion is
complete and the function $\Phi$ is monotonous, thus it has a fixed
point $\beta$. We let $\hat{N} = \beta$.

This way we can interpret every proposition $A$ that does not
contain Skolem symbols. Finally, we interpret the symbols $f_{x_1,
\ldots, x_p, y_1, \ldots, y_n, A}$ as the functions mapping $a_1, \ldots,
a_p$ to the function mapping $b_1, \ldots, b_n$ to $\llbracket A
 \rrbracket_{a_1/x_1, \ldots, a_p/x_p,b_1/y_1, \ldots, b_n/y_n}$.}

\section{Expressivity}
\label{secexpress}

We shall now see that, despite the non-determinism of the reduction,
given in Definition \ref{defreddep},
the uniqueness of results may be guaranteed
for some schemes, and that every function that is
provably total in $HA_2$ can be expressed by such a scheme.

If $n$ is a natural number, we write $\underline{n}$ for the
term $S^{n}(0)$.

\begin{proposition}[Parigot's numerals \cite{Parigot}]
Let $n$ be a natural number, then there exists a scheme
$\rho_n$ of type $N(\underline{n})$.
\end{proposition}

\proof{Let $A = (0~\epsilon_1~c)$ and $B =
(\fa y~(N(y) \Rightarrow y~\epsilon_1~c \Rightarrow
S(y)~\epsilon_1~c))$. Take
$$\rho_0 = \Lambda c \lambda_A \lambda_B~\var{A}$$
and
$$\rho_{n+1} =
\Lambda c \lambda_A
\lambda_B~(\var{B}~\underline{n}~\rho_n~(\rho_n~c~\var{A}~\var{B}))$$}

\begin{proposition}[Witness property] 
Let $\ex x~A$ be a closed proposition.
From a scheme $t$ of type
$\ex x~A$ {\em i.e.}
$\fa c~((\fa x~(A \Rightarrow \epsilon_0(c))) \Rightarrow \epsilon_0(c))$
in the empty context, we can extract a term $b$ and a scheme of type
$[b/x]A$ in the empty context.
\end{proposition}

\proof{Consider a term variable $c$ of sort $\kappa_0$ and
$g = \var{\fa x~(A \Rightarrow \epsilon_0(c))}$.
The scheme $(t~c~g)$ has type $\epsilon_0(c)$ in the context
$\{\fa x~(A \Rightarrow \epsilon_0(c))\}$, thus its weak normal form
has the form
$(g~a~u)$ where $a$ is a term of sort $\iota$ and
$u$ a scheme of type $[a/x]A$ in the context
$\{\fa x~(A \Rightarrow \epsilon_0(c))\}$.
Let $e = f_{\ex x~A}$ and $w$ be a closed proof of
$\fa x~(A \Rightarrow \ex x~A)$.
Let $b = [e/c]a$ and
$v = \underline{[w/\fa x~(A \Rightarrow \ex x~A)]}_{\varnothing}[e/c]u$.
The scheme $v$ has type $[b/x]A$ in the empty context.}

\medskip

From the witness property we get the expressibility of all functions
that are provable in $HA_2$. We need first to use the following result
of elementary logic.

\begin{proposition}
For every computable function $f$ from ${\mathbb N}^n$ to
${\mathbb N}$, there exists a proposition $A$ such that
$[\underline{p}_1/x_1, \ldots,
\underline{p}_n/x_n,\underline{q}/y]A$ is provable in $HA_2$ if
and only if $q = f(p_1, \ldots, p_n)$.
\end{proposition}

\begin{definition}[Provably total function]
The function $f$ is said to be
{\em provably total} in $HA_2$ if
$$\fa x_1~(N(x_1) \Rightarrow \ldots
\Rightarrow \fa x_n~(N(x_n) \Rightarrow \ex y~(N(y) \wedge
A))\ldots)$$
is provable in $HA_2$.
\end{definition}

\begin{theorem}
For every computable function $f$ provably total in $HA_2$,
there exists a scheme $t$ such that for all
$p_1, ..., p_n$, the normal form of the witness extracted from the
scheme
$(t~\underline{p}_1~\rho_{p_1}~\underline{p}_2~\rho_{p_2}~\ldots~
\underline{p}_n~\rho_{p_n})$
is $\underline{f(p_1, ..., p_n)}$.
\end{theorem}

\proof{Take any scheme of type
$\fa x_1~(N(x_1) \Rightarrow \ldots
\Rightarrow \fa x_n~(N(x_n) \Rightarrow \ex y~(N(y) \wedge
A))\ldots)$.}

Whether the set of functions provably total in HA$_2$ is equal or
a strict subset of the set of functions that can be expressed in 
the scheme calculus, is left as an open problem. 

\section{Future Work}

Besides $HA_2$, Theorem \ref{thnorm} applies to many theories
{\em e.g.} simple type theory and some variants of set theory.
When they cannot be defined in the theory, all connectives and
quantifiers must be taken as primitive, like in \cite{DW}. Although tedious,
the normalization proof generalizes smoothly.

A more challenging problem is to prove normalization for
other reduction strategies than weak minimal reduction. This probably
requires to generalize proofs
by reducibility to cases where reduction and substitution do not commute.

\section{Acknowledgments}

The authors want to thank the anonymous referees who helped them 
to improve the paper in many respects.
This work is partially supported by NSFC 60673045, NSFC 60833001
and NSFC 60721061.

\bibliographystyle{elsarticle-num}
\bibliography{schemes}

\begin{thebibliography}{10}
\expandafter\ifx\csname url\endcsname\relax
  \def\url#1{\texttt{#1}}\fi
\expandafter\ifx\csname urlprefix\endcsname\relax\def\urlprefix{URL }\fi
\expandafter\ifx\csname href\endcsname\relax
  \def\href#1#2{#2} \def\path#1{#1}\fi

\bibitem{Kleene}
S.~C. Kleene, Introduction to Metamathematics, North-Holland, 1952.

\bibitem{Zaionc}
M.~Zaionc, Probabilistic approach to the lambda definability for fourth order
  types, Electronic Notes in Theoretical Computer Science 140 (2005) 41--54.

\bibitem{Joly}
T.~Joly, On lambda-definability {I}: the fixed model problem and
  generalizations of the matching problem, Fundamenta Informaticae 65~(1-2)
  (2005) 135--151.

\bibitem{contfree}
G.~Dowek, Y.~Jiang, Enumerating proofs of positive formulae, The Computer
  Journal 52~(7) (2009) 799--807.

\bibitem{BenYelles}
C.~B. Ben-Yelles, Type-assignment in the lambda-calculus; syntax and semantics,
  Ph.D. thesis, University Coll. of Swansea (1979).

\bibitem{Zaionc1988}
M.~Zaionc, Mechanical procedure for proof construction via closed terms in
  typed lambda-calculus, Journal of Automated Reasoning 4 (1988) 173--190.

\bibitem{TakahashiAkamaHirokawa}
M.~Takahashi, Y.~Akama, S.~Hirokawa, Normal schemes and their grammar,
  Information and Computation 152~(2) (1996) 144--153.

\bibitem{ComonJurski}
H.~Comon, Y.~Jurski, Higher-order matching and tree automata, in: Computer
  Science Logic, 1997, pp. 157--176.

\bibitem{BrodaDamas}
S.~Broda, L.~Damas, A context-free grammar representation for normal
  inhabitants of types in {TA}-lambda., in: EPIA'01, Vol. 2258 of Lecture Notes
  in Artificial Intelligence, Springer-Verlag, 2001, pp. 145--159.

\bibitem{BrodaDamas2}
S.~Broda, L.~Damas, On long normal inhabitants of a type, Journal of Logic and
  Computation 15 (2005) 353--390.

\bibitem{bracket}
G.~Dowek, Y.~Jiang, Eigenvariables, bracketing and the decidability of positive
  minimal predicate logic, Theoretical Computer Science 360 (2006) 193--208.

\bibitem{Salvati}
S.~Salvati, Recognizability in the simply typed lambda-calculus, in: Wollic,
  2009.

\bibitem{Boudol}
G.~Boudol, The lambda-calculus with multiplicities, Tech. Rep. 2025, Institut
  National de Recheche en Informatique et en Automatique (1993).

\bibitem{TroelstraSchwichtenberg}
A.~Troelstra, H.~Schwichtenberg, Basic Proof theory, Cambridge University
  Press, 1996, 2000.

\bibitem{Leivant}
D.~Leivant, Assumption classes in natural deduction, Zeitschrift f\"ur
  mathematische Logik und Grundlagen der Mathematik 25 (1979) 1--4.

\bibitem{DHK}
G.~Dowek, T.~Hardin, C.~Kirchner, Theorem proving modulo, Journal of Automated
  Reasoning 31 (2003) 33--72.

\bibitem{DWPeano}
G.~Dowek, B.~Werner, Arithmetic as a theory modulo, in: J.~Giesel (Ed.), Term
  rewriting and applications, Vol. 3467 of Lecture Notes in Computer Science,
  Springer-Verlag, 2005, pp. 423--437.

\bibitem{DW}
G.~Dowek, B.~Werner, Proof normalization modulo, The Journal of Symbolic Logic
  68 (2003) 1289--1316.

\bibitem{Girard}
J.-Y. Girard, Une extension de l'interpr\'etation de {G}\"odel \`a l'analyse,
  et son application \`a l'\'elimination des coupures dans l'analyse et la
  th\'eorie des types, in: J.~Fenstad (Ed.), Second Scandinavian Logic
  Symposium, Vol.~63 of Studies in Logic and the Foundations of Mathematics,
  North-Holland, 1971, pp. 62--92.

\bibitem{Parigot}
M.~Parigot, Programming with proofs: A second order type theory, in:
  H.~Ganzinger (Ed.), European Symposium on Programming, Vol. 300 of Lecture
  Notes in Computer Science, Springer-Verlag, 1988, pp. 145--159.

\end{thebibliography}

\end{document}